\documentclass[runningheads]{llncs}

\usepackage[misc,geometry]{ifsym}
\usepackage{color, colortbl}
\usepackage[english]{babel}
\usepackage{changepage}
\usepackage{mathtools}
\usepackage{enumerate}
\usepackage{hyperref}
\usepackage{graphicx}
\usepackage{multirow}
\usepackage{tabularx}
\usepackage{booktabs}
\usepackage{amsmath}
\usepackage{amssymb}
\usepackage{float}
\usepackage{calc}
\usepackage{tcolorbox}

\tcbset{colback=pink}

\renewenvironment{quote}{%
   \list{}{%
     \leftmargin0.5cm   % this is the adjusting screw
     \rightmargin\leftmargin
   }
   \item\relax
}
{\endlist}

\begin{document}

\title{Disparate Impact Diminishes Consumer Trust Even for Advantaged Users}

\titlerunning{Disparate Impact Diminishes Consumer Trust}

\author{
Tim Draws\inst{1,2\thanks{current affiliation}}\textsuperscript{(\Letter)}%\orcidID{0000-0001-5053-4674}
\and
Zolt\'{a}n Szl\'{a}vik\inst{1,3\textsuperscript{$\star$}}%\orcidID{0000-0002-2781-3795}
\and
Benjamin Timmermans\inst{1,4\textsuperscript{$\star$}}%\orcidID{0000-0002-5130-4309}
\and
Nava Tintarev\inst{5}%\orcidID{0000-0002-5007-5161}
\and
Kush R. Varshney\inst{4}%\orcidID{0000-0002-7376-5536}
\and
Michael Hind\inst{4}%\orcidID{0000-0002-7247-7225}
}

\authorrunning{Draws et al.}

\institute{
IBM Center for Advanced Studies Benelux\\
\and
Delft University of Technology\\
\email{\href{mailto:t.a.draws@tudelft.nl}{t.a.draws@tudelft.nl}}
\and
myTomorrows\\
\email{\href{mailto:zoltan.szlavik@mytomorrows.com}{zoltan.szlavik@mytomorrows.com}}
\and
IBM Research\\
\email{\href{mailto:b.timmermans@nl.ibm.com}{b.timmermans@nl.ibm.com}, \href{mailto:krvarshn@us.ibm.com}{krvarshn@us.ibm.com}, \href{mailto:hindm@us.ibm.com}{hindm@us.ibm.com}}
\and
Maastricht University\\
\email{\href{mailto:n.tintarev@maastrichtuniversity.nl}{n.tintarev@maastrichtuniversity.nl}}
}

\maketitle

% \blfootnote{Corresponding author: Tim Draws; \email{t.a.draws@tudelft.nl}}

\begin{tcolorbox}
This is an unedited manuscript accepted for presentation at the \textit{16th International Conference on Persuasive Technologies} in 2021. The published version of the paper is available at \url{https://doi.org/10.1007/978-3-030-79460-6_11}.
\end{tcolorbox}

\begin{abstract}
  Systems aiming to aid consumers in their decision-making (e.g., by implementing persuasive techniques) are more likely to be effective when consumers trust them. However, recent research has demonstrated that the machine learning algorithms that often underlie such technology can act unfairly towards specific groups (e.g., by making more favorable predictions for men than for women). An undesired disparate impact resulting from this kind of algorithmic unfairness could diminish consumer trust and thereby undermine the purpose of the system. We studied this effect by conducting a between-subjects user study investigating how (gender-related) disparate impact affected consumer trust in an app designed to improve consumers' financial decision-making. Our results show that disparate impact decreased consumers' trust in the system and made them less likely to use it. Moreover, we find that trust was affected to the same degree across consumer groups (i.e., advantaged and disadvantaged users) despite both of these consumer groups recognizing their respective levels of personal benefit. Our findings highlight the importance of fairness in consumer-oriented artificial intelligence systems.
  
\end{abstract}

%---------- Beginning text ----------%

\section{Introduction}\label{sec:intro}

Applications that seek to advise or nudge consumers into better decision-making (e.g., concerning personal health or finance) %may pursue their aims by implementing persuasive technology (PT)~\cite{Oinas-Kukkonen2008} but 
can only be effective when consumers trust their guidance. Trustworthiness is an essential aspect in the design of such \textit{persuasive technology} (PT); i.e., technology aiming to change attitudes or behaviors without using coercion or deception~\cite{Nickel2012,Oinas-Kukkonen2008,Verbeek2006}; because consumers are unlikely to use (or be persuaded by) systems that they do not trust~\cite{Muir1996,Sattarov2019}. Recent research has identified several factors that affect consumer trust in this context; including consumers' emotional states~\cite{Ahmad2018} as well as the system's reliability~\cite{Nickel2012} and transparency~\cite{Sattarov2019}. Moreover, it has been argued that trust also depends on \textit{moral expectations} that consumers have towards the technology they use~\cite{Nickel2012,Sattarov2019}. Consumer trust could increasingly depend on such moral expectations as more systems implement machine learning algorithms (e.g., in personal health~\cite{Purpura2011,Sattarov2019} or finance~\cite{Lieber2014} applications) that make them harder to scrutinize.

A specific moral expectation that acts as a requirement for trust in this context may be \textit{fairness}~\cite{Varshney2019}. When nudges and advice are tailored to the individual consumer using machine learning, consumers may expect that the system acts fairly towards different consumer groups (e.g., concerning race or gender). Nudging or advising such that the degree of positive impact that the system has on consumers' lives \textit{varies} with group membership could constitute an undesired \textit{disparate impact}. For example, a \textit{robo-advisor} (i.e., PT designed to improve consumers' financial situation~\cite{Lieber2014}) could have a disparate impact by systematically recommending ``safer'', lower-risk investments to female consumers compared to male consumers, yielding them lower returns. Such disparate impact would violate the moral expectation of fairness and thereby undermine consumer trust.

Employing machine learning in consumer-oriented applications often holds the promise of increasing their usefulness to the individual consumer~\cite{Purpura2011,Yang2018a} but also bears a greater vulnerability for disparate impact. Recent research has demonstrated that machine learning algorithms may unfairly discriminate based on group membership~\cite{angwin2019machine,Barocas2016,Ntoutsi2020}. Such discrimination is referred to as \emph{algorithmic unfairness} if a pre-defined notion of fairness is violated~\cite{Ntoutsi2020,Verma2018} and can easily lead to an undesired disparate impact~\cite{Barocas2016,Feldman2015}. For example, outcomes in advice from robo-advisors may differ between groups, given that financial advice has historically been gender-biased to the disadvantage of female consumers~\cite{Baeckstrom2018,Mullainathan2012} and algorithmic unfairness often results from disparities in the historical data that is used to train the algorithm~\cite{Ntoutsi2020}. Although several methods have been developed to mitigate algorithmic unfairness~\cite{Bellamy2019}, in many cases it is currently not possible to do so to a satisfactory degree~\cite{Corbett-Davies2018}.

Disparate impact is thus a realistic issue that could undermine the efficacy of consumer-oriented artificial intelligence (AI) systems. It has been argued that fairness plays a key role in fostering trust in AI~\cite{Arnold2019,Rossi2019,Toreini2020,Varshney2019,Varshney2020}. However, to the best of our knowledge, no previous work has \textit{studied} the influence of undesired disparate impact (i.e., as a result of algorithmic unfairness) on consumer trust. It is further unclear whether unfairly advantaged consumers are affected to the same degree as disadvantaged consumers in this context. That is, the influence of disparate impact on consumer trust may depend on \textit{perceived personal benefit} (i.e., advantaged users trusting the system's advice despite disparate impact as long as they personally benefit) or not (i.e., advantaged users losing trust in lockstep with disadvantaged users despite a perceived personal benefit). We study the effect of disparate impact on consumer trust at the use case of gender bias in robo-advisors by investigating the following research questions:

\begin{itemize}
    \item \textbf{RQ1.} Does an apparent disparate impact of a robo-advisor affect the degree of trust that consumers place in it?
    \item \textbf{RQ2.} Does disparate impact affect the trust of unfairly advantaged consumers to a different degree than that of unfairly disadvantaged consumers?
\end{itemize}

% \paragraph{\textbf{RQ1.}} Does an apparent disparate impact of a robo-advisor affect the degree of trust that consumers place in it?

% \paragraph{\textbf{RQ2.}} Does disparate impact affect the trust of unfairly advantaged consumers to a different degree than that of unfairly disadvantaged consumers?

To answer these questions, we conducted a between-subjects user study where we exposed participants to varying degrees of disparate impact of a robo-advisor (i.e., advantaging male users; see Section~\ref{sec:method}). Our results show that disparate impact negatively affected consumer's trust in the robo-advisor and decreased their willingness to use it (see Section~\ref{sec:results}). Furthermore, we find that, despite both groups recognizing their respective personal (dis)advantage, \emph{both} the disadvantaged group (women) as well as the advantaged group (men) experienced the same decrease in trust when learned about a disparate impact of the robo-advisor. Our findings underline the importance of ensuring algorithmic fairness in consumer-oriented (AI) systems when aiming to maintain consumer trust.

\section{Background and Related Work}\label{sec:relWork}

We study the effect of disparate impact on consumer trust at the use case of gender bias in robo-advisors. Our reasons for choosing the financial domain here are threefold. First, algorithmic decision-making is already widespread in consumer-oriented financial applications (e.g., in robo-advisors)~\cite{Lieber2014}. Second, algorithmic decision-making in such systems is highly impactful: it directly affects consumers' financial situations and thereby their life quality. Third, (human) financial advice has traditionally been gender-biased, underestimating and disadvantaging female consumers~\cite{Mullainathan2012,Baeckstrom2018}. Historical data on financial advice thus contain these biases. If the algorithms that underlie robo-advisors are trained using these data, robo-advisors may have according disparate impact.

\textit{Trust in AI systems.} Consumers do not use systems that they do not trust \cite{Muir1996}. That is why trust is an important aspect in the interaction between consumer-oriented AI systems (e.g., those implementing PT) and consumers~\cite{Ahmad2018,Nickel2012,Sattarov2019,Verbeek2006}. Recent research has linked trust in such systems to the reliability~\cite{Nickel2012} and transparency~\cite{Sattarov2019} of the system at hand as well as consumers' emotional states~\cite{Ahmad2018} and moral expectations~\cite{Nickel2012,Sattarov2019}. Such moral expectations may gain in importance as systems increasingly rely on machine learning algorithms~\cite{Lieber2014,Orji2018,Purpura2011,Sattarov2019,Yang2018a}. Moreover, whereas in some cases consumers fall prey to \textit{automation bias} (i.e., a tendency to prefer automated over human decisions)~\cite{Cummings2004}, in other cases, they experience what has been referred to as \emph{algorithm aversion}: a tendency to prefer human over algorithmic advice \cite{Diab2011,Onkal2009,Promberger2006}. Research has shown that algorithm aversion can be the result of witnessing how an algorithm errs~\cite{Dietvorst2015}. Especially in cases where a machine learning algorithm acted \textit{unfairly}, leading to an undesired disparate impact (i.e., violating consumers' moral expectations and reflecting erroneous decision-making), consumer trust could thus be diminished.

\textit{Measuring and mitigating algorithmic unfairness.} Research has demonstrated that machine learning algorithms can make biased (unfair) predictions to the disadvantage of specific groups~\cite{angwin2019machine,Barocas2016,Ntoutsi2020,vigdor2019apple}. For instance, AI systems may discriminate between white and black defendants in predicting their likelihood of re-offending~\cite{angwin2019machine} and between male and female consumers in predicting their creditworthiness~\cite{vigdor2019apple}. Several methods have been proposed to measure and mitigate biases in algorithmic decision-making~\cite{Bellamy2019,Hardt2016,Mary2019,Mehrabi2019,Ntoutsi2020,Zafar2017}. Despite these efforts, the measurement and mitigation of algorithmic bias remain challenging~\cite{Corbett-Davies2018,Ntoutsi2020}.

\textit{Disparate impact and trust.} Algorithmic fairness has been identified as a core building block of trustworthy AI systems~\cite{Arnold2019,Rossi2019,Varshney2019,Toreini2020,Varshney2020}, yet few studies directly investigate the relationship between algorithmic fairness (or disparate impact) and consumer trust. %Some research nevertheless indicates that disparate impact as a result of algorithmic unfairness could have a dramatic impact on consumer trust. For instance, p
Participants in one study reported that learning about algorithmic unfairness induced negative feelings and that it might cause them to lose trust in a company or product~\cite{Woodruff2018}. Consumers have further expressed general concerns about disparate impact of AI on a societal level~\cite{Araujo2020} and are more likely to judge decisions as less fair and trustworthy if they are made by an algorithm as opposed to a human~\cite{Lee2018}. However, it has also been shown that the degree to which people are concerned about disparate impact depends on their personal biases~\cite{Otterbacher2018,Smith2020}. %Toreini et al. \cite{Toreini2020} propose a framework for the design and implementation of AI systems that aligns consumer trust with trustworthy (i.e., fair, explainable, auditable, and safe) machine learning. 
What remains unclear is to what extent disparate impact (as a result of algorithmic unfairness) affects consumer trust and, if so, who (i.e., unfairly advantaged and disadvantaged consumers) are affected in particular.

\section{Method}\label{sec:method}

To investigate the two research questions identified in Section~\ref{sec:intro}, we conducted a between-subjects user study. The setting of this study was a fictional scenario in which a bank offers a robo-advisor -- called the \emph{AI Advisor} -- to its customers. We aimed to perform a granular analysis of the effect of disparate impact on consumer trust by exposing participants to different degrees of disparate impact supposedly caused by the \emph{AI Advisor} and measuring their attitudes towards this system. Specifically, we analyzed whether the different degrees of disparate impact affected participant's \textit{trust} (i.e., whether they believed that the AI Advisor would make correct predictions and therefore benefit its users). To differentiate between this general notion of trust and related attitudes, we also measured \textit{willingness to use} and \textit{perceived personal benefit} concerning the \textit{AI Advisor}.

\subsection{Operationalization}\label{sec:operationalization}

\paragraph{\textbf{Dependent Variables.}} Our experiment involved measuring participants' attitudes towards the \emph{AI advisor}; specifically \emph{trust}, \textit{willingness to use}, \emph{perceived personal benefit}. % We measured each of these constructs on 7-point Likert scales ranging from ``strongly disagree'' to ``strongly agree'' and coded the points from the Likert scale as integers ranging from -3 (strongly disagree) to 3 (strongly agree). 
Each variable was measured twice: once after participants saw general user statistics (Step 1; see Section \ref{sec:procedure}) and once after participants saw gender-specific user statistics on the \emph{AI advisor} (Step 2). We computed difference scores from these two measurements that reflected how seeing the gender-specific statistics affected participant's attitudes as compared to their baseline attitudes.
% how participants' attitudes had changed after seeing the gender-specific statistics. Crucially, this incorporated participant's baseline attitudes into the variables we analyzed.

\begin{itemize}
    \item \emph{Change in Trust (Continuous).} Participants rated their trust by responding to the item ``In general, the AI advisor can be trusted to make correct recommendations'' on a 7-point Likert scale. We coded all responses on an ordinal scale ranging from $-3$ (strongly disagree) to $3$ (strongly agree) and subtracted the second measurement from the first to compute the \emph{change in trust}. Values could thus range from $-6$ to $6$.%\nt{There are many notions of trust, and somewhere there needs to be a discussion of why perception of system accuracy is the most relevant to study.}
    \item \emph{Change in Willingness to Use (Categorical).} Participants could respond to the item ``I would personally use the AI Advisor'' with either ``yes'' or ``no''. We recorded whether their answer had changed (i.e., ``yes'' to ``no'' or vice versa) or stayed the same in the second measurement. This variable thus encompassed three categories.
    \item \emph{Change in Perceived Personal Benefit (Continuous).} Participants rated their perceived personal benefit by responding to the item ``I would personally benefit from using the AI advisor'' on a 7-point Likert scale. To compute the \emph{change in perceived personal benefit}, we again subtracted the second measurement from the first. Values could thus range from $-6$ to $6$.
\end{itemize}

\paragraph{\textbf{Independent Variable.}} Our experiment varied depending on the condition that a participant was placed in (see Section \ref{sec:procedure}):

\begin{itemize}
    \item \emph{Condition.} During the experiment, we showed participants a table with user statistics of bank customers that use the \emph{AI advisor}. These statistics, supposedly showing the average change in bank account balance for users and non-users of the \textit{AI Advisor}, split by gender, differed depending on the condition a participant had been placed in. Each participant saw only one of four conditions: the \emph{control} condition (in which the statistics were balanced across genders, reflecting an absence of disparate impact) or one of three experimental conditions -- which we call \emph{little bias}, \emph{strong bias}, and \emph{extreme bias} -- that reflected varying degrees of disparate impact in favor of male consumers. Specifically, these different degrees of disparate impact represented scenarios in which female users of the \textit{AI advisor} were disadvantaged but still benefited from using the \textit{AI advisor} (little bias), did not benefit from the \textit{AI advisor} (strong bias), or would in fact benefit from \textit{not} using the \textit{AI advisor} (extreme bias). Table \ref{tbl:conditions} shows the numbers that were shown in the second statistics table in each of the conditions.
\end{itemize}

% \begin{table}
% 	\centering
% 	\caption{User statistics shown to participants in the different conditions.}\label{tbl:conditions}
% 	\begin{tabular}{lrrr}
% 		\toprule
% 		Condition & Male users & Female users & All Users\\
% 		\midrule
% 		Control & 20\% & 20\% & 20\%\\
% 		Little Bias & 25\% & 15\% & 20\%\\
% 		Strong Bias & 30\% & 10\% & 20\%\\
% 		Extreme Bias & 35\% & 5\% & 20\%\\
% 		\bottomrule
% 	\end{tabular} 
% \end{table}

\begin{table}
    \setlength{\tabcolsep}{7pt} % Default value: 6pt
    \renewcommand{\arraystretch}{1} % Default value: 1
	\centering
	\caption{Fictional gender-specific statistics shown to participants during the second step of the study across. Only the top left two cells (concerning users of the \emph{AI Advisor}) differed across conditions, reflecting varying degrees of disparate impact.}\label{tbl:conditions}
	\scalebox{0.9}{
	\begin{tabular}{l|ll}
		\toprule
		 & Using & Not Using\\ & AI Advisor & AI Advisor\\
		\midrule
		Male & 20\% & 10\%\\
		Female & 20\% & 10\%\\
		\midrule
		All & 20\% & 10\%\\
		\bottomrule
		\multicolumn{3}{c}{{\small \textit{Control} condition} \vspace{0.5em}}
	\end{tabular}
	}
	\quad
    \scalebox{0.9}{
	\begin{tabular}{l|ll}
		\toprule
		 & Using & Not Using\\ & AI Advisor & AI Advisor\\
		\midrule
		Male & 25\% & 10\%\\
		Female & 15\% & 10\%\\
		\midrule
		All & 20\% & 10\%\\
		\bottomrule
		\multicolumn{3}{c}{{\small \textit{Little bias} condition} \vspace{0.5em}}
	\end{tabular}
	}
	\quad
	\scalebox{0.9}{
	\begin{tabular}{l|ll}
		\toprule
		 & Using & Not Using\\ & AI Advisor & AI Advisor\\
		\midrule
		Male & 30\% & 10\%\\
		Female & 10\% & 10\%\\
		\midrule
		All & 20\% & 10\%\\
		\bottomrule
		\multicolumn{3}{c}{{\small \textit{Strong bias} condition}}
	\end{tabular}
	}
	\quad
	\scalebox{0.9}{
	\begin{tabular}{l|ll}
		\toprule
		 & Using & Not Using\\ & AI Advisor & AI Advisor\\
		\midrule
		Male & 35\% & 10\%\\
		Female & 5\% & 10\%\\
		\midrule
		All & 20\% & 10\%\\
		\bottomrule
		\multicolumn{3}{c}{{\small \textit{Extreme bias} condition}}
	\end{tabular}
	}
\end{table}

\paragraph{\textbf{Individual Differences and Descriptive Statistics.}}

We took two additional measurements to enable more fine-grained analyses and describe our sample:

\begin{itemize}
    \item \emph{Gender.} Participants could state which gender they identified with by picking from the options ``male'', ``female'', and ``other / not specified''.
    \item \emph{Age.} Participants could write their age in an open text field.
\end{itemize}

\subsection{Hypotheses}\label{sec:hypotheses}

Based on the research questions \textbf{RQ1} and \textbf{RQ2} introduced in Section \ref{sec:intro}, the related work from Section \ref{sec:relWork}, and the experimental setup described in this section, we formulated several hypotheses. We expected that disparate impact will decrease consumer trust (H1a) and that consumers will be less likely to use the \textit{AI Advisor} (H1b) if it has disparate impact (i.e., the stronger the disparate impact, the lower consumer trust and willingness to use the \textit{AI Advisor}). We predicted that disparate impact would affect the perceived personal benefit of male consumers differently compared to female consumers (i.e., following what the displayed statistics suggest; H2a). Accordingly, we further expected that the decrease in trust described in H1a would be moderated by gender (H2b). That is, we predicted that the trust of advantaged consumers (i.e., men) would be affected differently compared to disadvantaged consumers (i.e., women).

\begin{itemize}
    \item \textbf{H1a.} Consumers who are exposed to statistics that reveal a disparate impact of a robo-advisor in favor of male users will \textit{trust this system less} to give correct recommendations compared to consumers who are exposed to balanced statistics.
    \item \textbf{H1b.} Consumers who are exposed to statistics that reveal a disparate impact of a robo-advisor in favor of male users will be \textit{less likely to use this system} compared to consumers who are exposed to balanced statistics.
    \item \textbf{H2a.} The effect of statistics suggesting a disparate impact of a robo-advisor in favor of men on \textit{perceived personal benefit} is moderated by gender.
    \item \textbf{H2b.} The effect of statistics suggesting a disparate impact of a robo-advisor in favor of men on \textit{consumer trust} is moderated by gender.
\end{itemize}

\subsection{Procedure}\label{sec:procedure}

We set up our user study by creating a task on the online study platform \emph{Figure Eight}.\footnote{Since conducting this study in June 2019, \emph{Figure Eight} has been renamed to \emph{Appen}. More information can be found at \url{https://appen.com}.} %\nt{already introduced in the beginning of the section? debating is this should only be under Participants?}. 
% This platform offers requesters to publish tasks for data collection. Crowd workers can complete these tasks in exchange for a financial reward paid by the requesters.
Before commencing with the experiment, participants were shown a short introduction and asked to state their gender and age. The experiment consisted of two steps. Whereas Step 1 was the same for all participants, Step 2 differed depending on which one of four conditions a participant had been assigned to.

\paragraph{\textbf{Step 1.}} We introduced participants to a fictional scenario in which they could activate a robo-advisor -- called the \emph{AI advisor} -- in their banking app:

\vspace{-.5em}

\begin{quote}

    {\small \emph{``Imagine your bank offers a digital assistant called the `AI advisor'. If you activate the AI advisor in your banking app, it will monitor your financial situation and give you relevant recommendations that may improve your financial situation. For example, it may suggest saving strategies or recommend investments.''}}
\end{quote}

\vspace{-.5em}

\noindent Additionally, to promote the idea that the AI Advisor is generally reliable, participants were given an idea of whether people benefit from using the \emph{AI Advisor}:

\vspace{-.5em}

\begin{quote}
	{\small \emph{``Overall statistics suggest that people benefit from using the AI advisor. The bank account balance of bank customers who use the AI advisor increases by an average of 20\% every year, whereas the balance of customers who don't use the AI advisor increases by an average of only 10\% per year.''}}
\end{quote}

\vspace{-.5em}

\noindent Below was a table displaying the mentioned statistics. We then measured trust, willingness to use, and perceived personal benefit concerning the \emph{AI Advisor}.

\paragraph{\textbf{Step 2.}} Participants were led to a new page for the second step of the experiment. Here we added some additional information on the AI advisor:

\vspace{-.5em}

\begin{quote}
	{\small \emph{``Next to general statistics on all bank customers, we can also look at how the AI advisor performs for subgroups of bank customers. Below you can see the change in bank account balance for men and women in particular.''}}
\end{quote}

\vspace{-.5em}

Below this text was a table similar to the table in Step 1, but with two added rows that showed the average change in bank account balance per year for men and women in particular (see Table \ref{tbl:conditions}). Whereas the statistics for all bank customers overall, as well as for men and women \emph{not} using the \emph{AI advisor} was the same in all conditions, the statistics for men and women \emph{using} the \emph{AI advisor} varied depending on the condition they were assigned to (see Section \ref{sec:operationalization}). Each Table \ref{tbl:conditions} shows the displayed statistics for male and female users per condition. We then again measured trust, willingness to use, and perceived personal benefit.

\vspace{-.5em}

\subsection{Statistical Analyses}\label{sec:statAnalysis}

\paragraph{\textbf{Testing H1a and H2b.}} To test whether there is an effect of disparate impact on consumer trust (H1a) that is moderated by gender (H2b), we conducted a classical ANOVA with \emph{condition} and \emph{gender} as between-subjects factors and \emph{change in trust} as the dependent variable. A significant main effect of \textit{condition} on \textit{change in trust} in this analysis would suggest that \emph{change in trust} differed between conditions (H1a). In this case, we would perform posthoc analyses to investigate the differences between the conditions in more detail. A significant interaction effect between \textit{condition} and \textit{gender} would suggest that the conditions had a different effect for the disadvantaged group (i.e. female participants) compared to the advantaged group (i.e., male participants; H2b).% We checked whether the data satisfied the assumptions of the ANOVA\footnote{We use the Shapiro-Wilk test to analyze whether the data are normally distributed an Levene's test to check for homogeneity of variances.} and would conduct a Kruskal-Wallis test as a non-parametric alternative to the classical ANOVA in case the assumptions were not met. 

We further conducted a Bayesian ANOVA according to the protocol proposed by van den Bergh et al.~\cite{VanDenBergh2020}. Bayesian hypothesis tests involve the computation of the \textit{Bayes factor}, a quantitative comparison of the predictive power of two competing statistical models~\cite{Wagenmakers2018}. The Bayes factor weighs the evidence provided by the data and thus allows for direct model comparison. Practically, comparing different models (i.e., including or excluding an interaction effect of \textit{condition} and \textit{gender}) this way allowed for a richer interpretation of our results. We performed the Bayesian ANOVA using the software JASP~\cite{JASP2020} with default settings. We computed \textit{Bayes Factors} (BFs) by comparing the models of interest to a \textit{null model}\footnote{The \textit{null model} in this procedure consisted of only an intercept.} and interpreted them according to the guidelines proposed by Lee and Wagenmakers~\cite{Lee2013}, who adopted them from Jeffreys~\cite{jeffreys1939}.% (e.g., BFs larger than 100 indicate extreme evidence).

\vspace{-.5em}

\paragraph{\textbf{Testing H1b.}} We tested whether disparate impact affected participants' willingness to use the \textit{AI Advisor} by conducting a chi-squared test between \textit{condition} and \textit{change in willingness to use}. A significant result in this analysis would suggest that the number of participants' who changed their willingness to use the \textit{AI Advisor} differed across conditions.

\vspace{-.5em}

\paragraph{\textbf{Testing H2a.}} We conducted another ANOVA with \emph{condition} and \emph{gender} as between-subjects factors and \emph{change in perceived personal benefit} as dependent variable to test whether \textit{gender} acted as a moderator here. A significant interaction effect in this analysis would indicate that this was the case.

\vspace{-.5em}

\paragraph{\textbf{Significance Threshold and Correction for Multiple Testing.}} In all classical analyses we conducted, we aimed for a type 1 error probability of no more than 0.05. % This usually means setting the significance threshold to 0.05 and regarding any result with a $p$-value that falls below this threshold as statistically significant. 
However, by conducting our planned analyses we automatically tested a total of seven hypotheses: three in each ANOVA (i.e., two main effects and one interaction) and one in the chi-squared test. This meant that the probability of committing a type 1 error rose considerably \cite{Cramer2016}.%\footnote{Although we were only interested in four of these seven hypotheses, we corrected our significance threshold as if we tested seven hypotheses because we did not preregister our study.} 
Therefore, we adjusted our significance threshold by applying a Bonferroni correction, where the desired type 1 error rate is divided by the number of hypotheses that are tested \cite{Napierala2012}. In our main analyses we thus handled a significance threshold of $\frac{0.05}{7} = 0.007$ and only regarded results as statistically significant if their $p$-value fell below this adjusted threshold. The same procedure was applied for posthoc analyses comparing each of the four conditions with each other as this meant conducting ${4\choose 2} = 6$ hypothesis tests (i.e., adjusting the threshold to $\frac{0.05}{6} = 0.008$).

\vspace{-.5em}

\subsection{Participants}\label{sec:participants}

We recruited 567 participants via the \emph{Figure Eight} pool of contributors (554) and direct contacts (13). %\nt{I find it strange to mix these two channels of recruiting participants...}.
Seventy-three participants were excluded from the study because they either filled at least one of the obligatory text fields with less than 10 characters, took less than 60 seconds to complete the task, or took more than 10 minutes to complete the task. Furthermore, we did not analyze data of five participants who stated ``other / not specified'' as their gender because our study involved a disparate impact between male and female consumers. 

After exclusion, 489 participants remained. Of those, 238 (49\%) were male and 251 (51\%) were female; with a mean age of $41.9$ (sd = $13.1$). Participants recruited by Figure Eight received \$0.10 as payment for participation. Random allocation to the four conditions resulted in 124, 121, 121, and 122 participants in the \textit{control}, \textit{little bias}, \textit{strong bias}, and \textit{extreme bias} conditions, respectively.

\vspace{-.5em}

\section{Results}\label{sec:results}

\begin{figure}
	\centering
	\begin{minipage}{.465\linewidth}
		\centering
		\includegraphics[width=.9\linewidth]{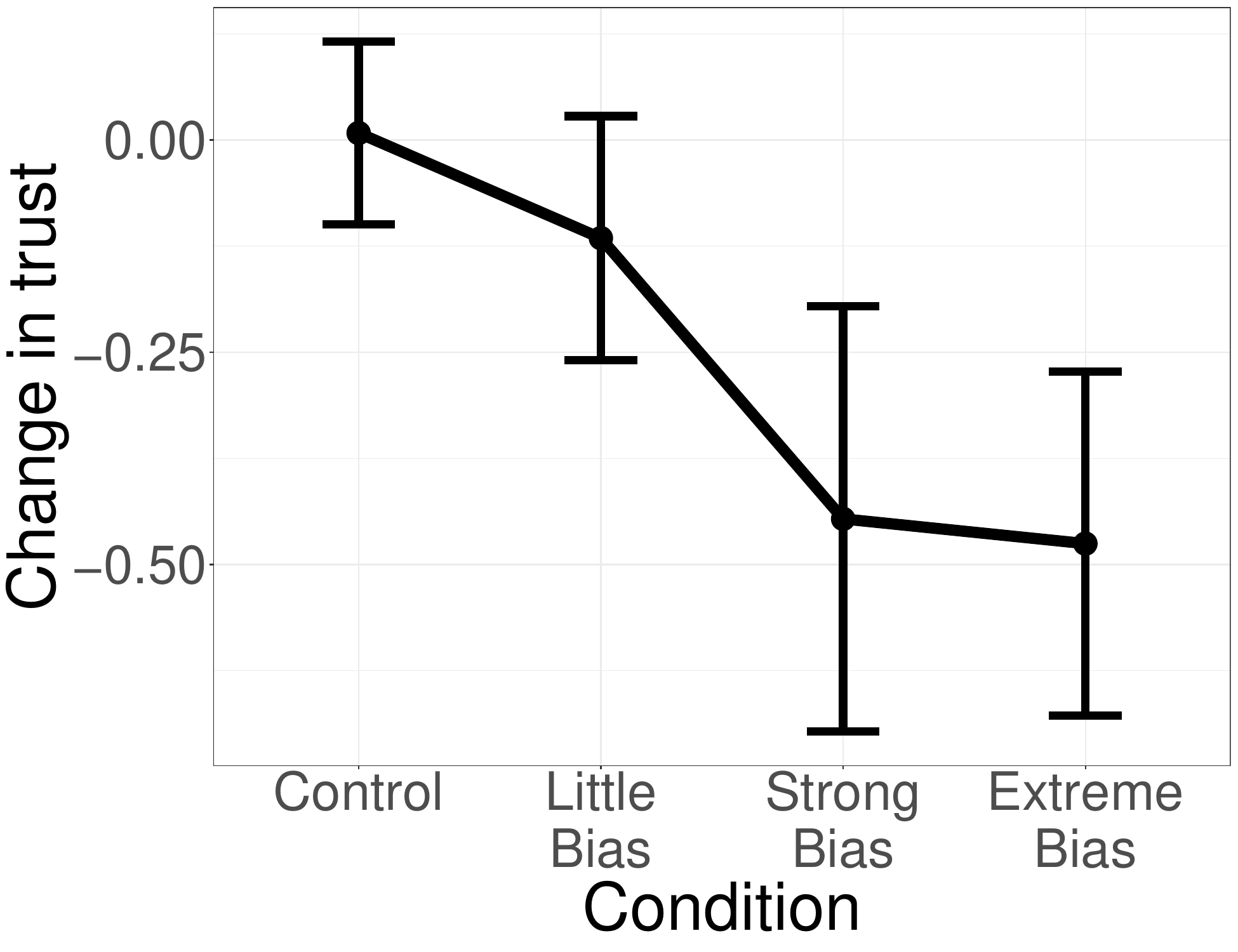}
	\end{minipage}%
	\begin{minipage}{.535\linewidth}
		\centering
		\includegraphics[width=.9\linewidth]{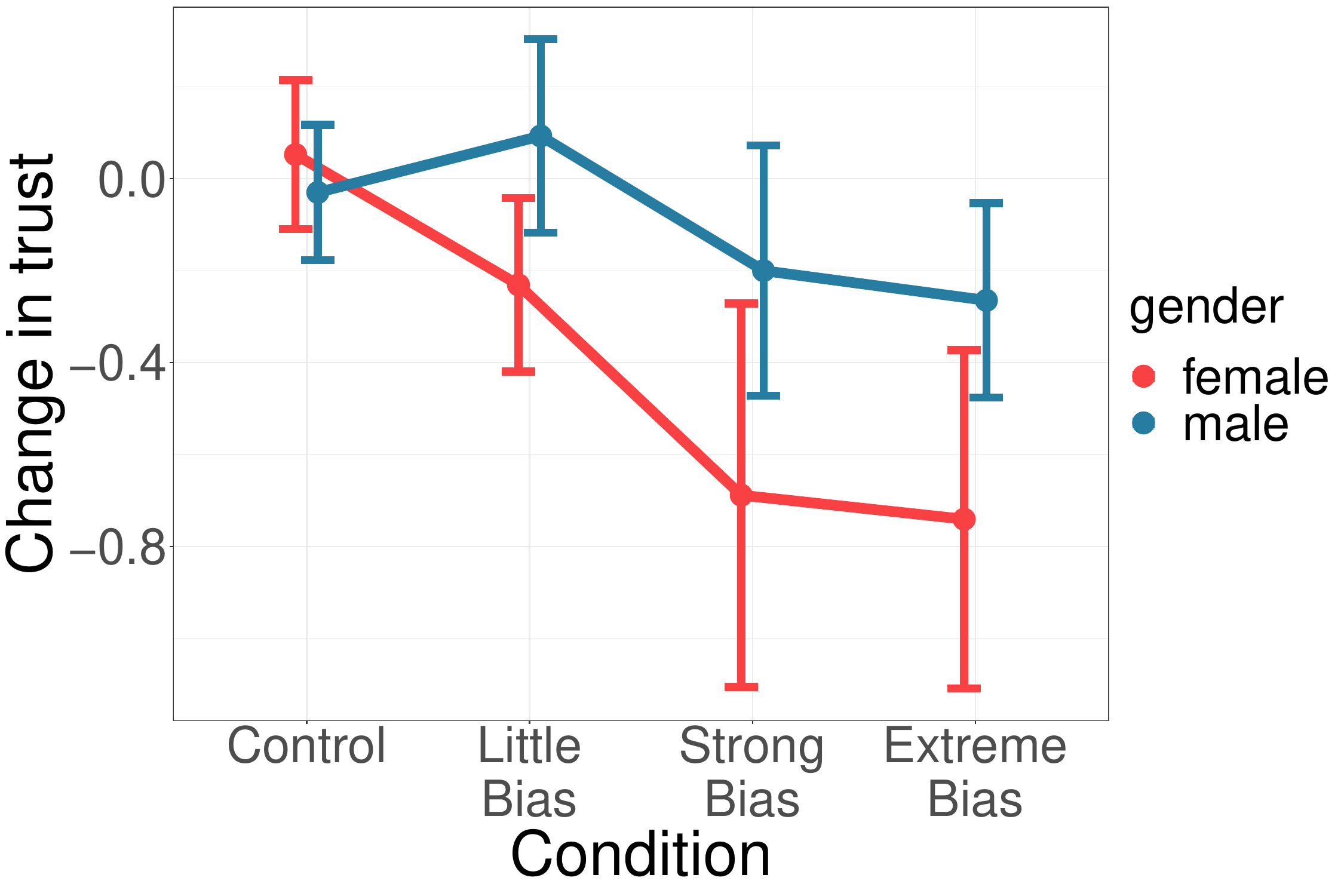}
	\end{minipage}
% 	\vspace*{5mm}
	\caption{\emph{Change in trust} across conditions for all participants (left-hand panel) and split by gender (right-hand panel). The error bars represent 95\% confidence intervals.}
	\label{fig:resultsTrust}
\end{figure}

\paragraph{\textbf{H1a: Disparate Impact Decreased Consumer Trust.}} As hypothesized, \emph{change in trust} differed across conditions (\emph{F} = 6.906, $p < 0.001$; see the left-hand panel of Figure \ref{fig:resultsTrust}). %However, a Shapiro-Wilk test showed that the data are not normally distributed ($W$ = 0.737, $p < 0.001$) and a Levene's test indicated that the assumption of equal variances was violated ($F$ = 6.882, $df$ = 7, $p < 0.001$). That is why we conducted a non-parametric alternative -- the Kruskal-Wallis test -- to confirm the ANOVA results. The Kruskal-Wallis test showed a significant main effect of condition ($\chi^2$ = 15.369, $df$ = 3, $p$ = 0.002). Additionally, we conducted a Bayesian ANOVA to quantify evidence in favor of this effect (see Table \ref{tbl:bayesianANOVA}). 
The results from the Bayesian ANOVA confirm this result, showing strong evidence for a main effect of condition ($\text{BF}_{10} = 70.02$, see Table \ref{tbl:bayesianANOVA}). To test for differences between the individual conditions, we conducted posthoc analyses (i.e., Mann-Whitney $U$ tests). Only the difference between the \emph{control} and \emph{extreme bias} conditions was significant ($W$ = 9368, $p < 0.001$). This suggests that participants lost trust due to disparate impact, but also that the unfairness needed to be comparatively extreme for this effect to occur.

\vspace{-1em}

\begin{table}[ht]
\setlength{\tabcolsep}{10pt} % Default value: 6pt
\renewcommand{\arraystretch}{1} % Default value: 1
\centering
  \caption{Bayesian ANOVA with \emph{change in trust} as dependent variable.}\label{tbl:bayesianANOVA}
  \scalebox{0.8}{\begin{tabular}{lrrrrr}
    \toprule
    Models & P(M) & P(M$\mid$Data) & BF$_{M}$ & BF$_{10}$ & Error \%  \\
    \midrule
    Null model & 0.200 & 8.209e-4 & 0.003 & 1.000 & \\
    condition & 0.200 & 0.057 & 0.244 & 70.024 & 0.001\\
    gender & 0.200 & 0.004 & 0.018 & 5.413 & 1.533e-6\\
    condition + gender & 0.200 & 0.735 & 11.116 & 895.772 & 1.638\\
    condition + gender +\\condition * gender & 0.200 & 0.202 & 1.012 & 245.894 & 1.978\\
  \bottomrule
\end{tabular}}
\end{table}

\vspace{-2em}

\vspace{-1em}

\paragraph{\textbf{H1b: Disparate Impact Decreased Willingness to Use.}} In accordance with disparate impact negatively affecting trust (H1a), it decreased participants' willingness to use the \textit{AI Advisor} (see Table \ref{tbl:willingnessUse}). The increasing proportion of participants who changed their attitude from ``yes'' to ``no'' as conditions reflected stronger disparate impact was statistically significant ($\chi^2 = 25.06$, $p < 0.001$).

\begin{table}[ht]
\setlength{\tabcolsep}{10pt} % Default value: 6pt
\renewcommand{\arraystretch}{1} % Default value: 1
\centering
  \caption{\textit{Change in willingness to use} across conditions. The labels $-$, $=$, and $+$ reflect changes from ``yes'' to ``no'', no change, ``no'' to ``yes'', respectively.}\label{tbl:willingnessUse}
  \scalebox{0.8}{\begin{tabular}{r|rrrr}
    \toprule
    & \multicolumn{4}{c}{\textbf{Condition}} \\
     \textbf{Change} & Control & Little Bias & Strong Bias & Extreme Bias \\
    \midrule
    $-$ & 1 & 9 & 23 & 16 \\
    $=$ & 121 & 111 & 97 & 105 \\
    $+$ & 2 & 1 & 1 & 1 \\
    \midrule
    total & 124 & 121 & 121 & 122 \\
  \bottomrule
\end{tabular}}
\end{table}

\begin{figure}
    \centering
    \includegraphics[width=.535\linewidth]{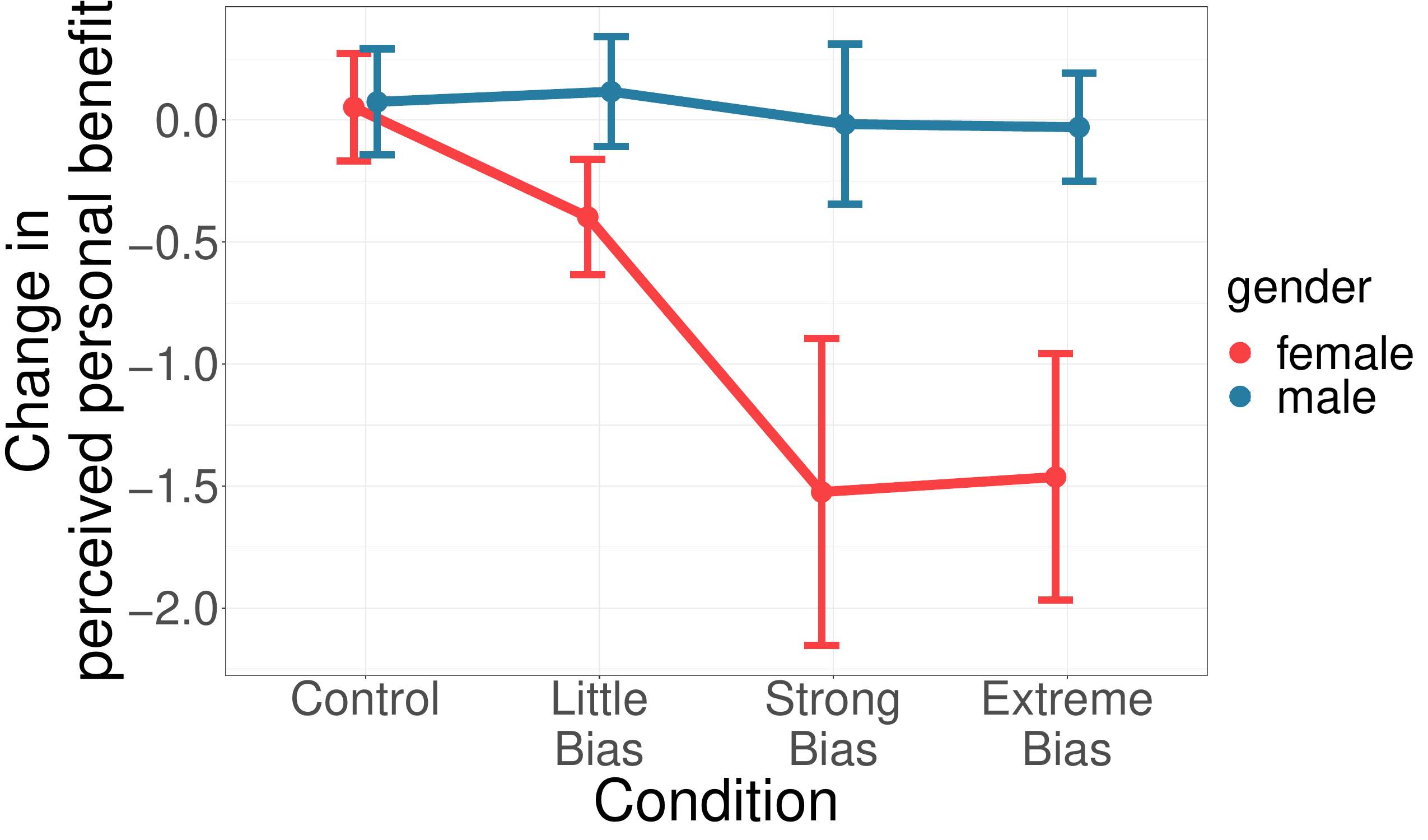}
    \caption{\emph{Change in perceived personal benefit} across conditions and split by gender. The error bars represent 95\% confidence intervals.}
    \label{fig:resultsBenefit}
\end{figure}

\paragraph{\textbf{H2a: Gender Moderated the Effect of Disparate Impact on Perceived Personal Benefit.}} As expected, the results from the second ANOVA show a significant interaction effect of \emph{condition} and \emph{gender} on \emph{change in perceived personal benefit} (\emph{F} = 8.525, $p < 0.001$). %A Bayesian ANOVA confirms this result, with the best model including both main effects for \emph{condition} and \emph{gender} as well as the interaction effect ($\text{BF}_{10}$ = 3.556798e+14). 
% We thus find that, although male participants experienced the same decrease in trust as female participants as a result of the apparent disparate impact (see results on \textbf{H2}), there \emph{was} a gender difference in terms of perceived personal benefit across conditions. In other words, men recognized their advantage but nevertheless deemed unfair versions of the \emph{AI advisor} as less trustworthy.
This means that male participants' perceived personal benefit was affected differently compared to that of female participants. More specifically, Figure \ref{fig:resultsBenefit} shows that whereas men's perceived personal benefit did not change due to seeing the gender-specific user statistics across conditions, female participants perceived increasingly lower levels of personal benefit as disparate impact (to their disadvantage) became more severe.

\paragraph{\textbf{H2b: Male Consumers Experienced the Same Decrease in Trust as Female Consumers.}} In contrast to what we hypothesized, we do not find a significant interaction effect of \emph{condition} and \emph{gender} on \textit{change in trust} ($F = 2.094$, $p = 0.096$; see the right-hand panel of Figure \ref{fig:resultsTrust}). We can therefore not conclude that the conditions had a different effect on male participant's \emph{change in trust} compared to that of female participants. The Bayesian ANOVA confirms this result: the model containing just two main effects for condition and gender explain the data best ($\text{BF}_{10}$ = 895.77; see Table \ref{tbl:bayesianANOVA}) and roughly four times better than the model that includes the interaction effect ($\text{BF}_{10}$ = 245.89). This suggests that unfairly advantaged and disadvantaged participants (i.e., men and women, respectively) experienced the same decrease in trust due to algorithmic unfairness despite diverging levels of perceived personal benefit (H2a).

\section{Discussion}

In this paper, we presented a between-subjects user study that aimed to investigate the influence of algorithmically-driven disparate impact on consumer trust at the use case of gender-bias in robo-advisors. Our results suggest that disparate impact -- at least when it is extreme -- decreases trust and makes consumers less likely to use such systems. We further find that, although disadvantaged and advantaged users recognize their respective levels of personal benefit in scenarios of disparate impact, both experience equally decreasing levels of trust when they learn about a disparate impact caused by the system at hand. Our work contributes to a growing body of literature that highlights the importance of ensuring fairness and avoiding disparate impact of consumer-oriented AI systems.

\subsection{Implications}
% \nt{May be that fairness is not possible, should one then not have a system at all? Perhaps here (richer) explanations can bridge the gap between accountability of the companies and the reliance on system advice by consumers}
Our findings have implications for consumers as well as industry. Consumers should be aware that machine-learning-based applications can be biased. If disparate impact is an important factor for consumer trust, consumers need to think critically when using such systems. One potential way forward for consumers would be to demand from companies to publish independently carried out research into the (algorithmic) fairness and impact of their products.

Publishers of consumer-oriented AI systems need to establish algorithmic fairness in their products and avoid disparate impact to serve consumers effectively. Our findings show that failing to do so may lead to a decrease in consumers' trust and willingness to use such systems.

\subsection{Limitations and Future Work}

Our study is subject to at least five important limitations. First, we studied the effect of disparate impact on consumer trust at a specific use case: a binary gender bias in robo-advisors. This makes our results difficult to generalize because many other forms of bias (including those based on race, religion, or sexual orientation) as well as other AI systems (e.g., for recommendations of medical treatment, tourist attractions, or movies) exist. It is easy to imagine how consumer trust could be affected differently when, for example, disparate impact concerns small minorities, multitudes of gender identities (or another consumer characteristic), a chosen group membership such as consumers' profession, or a system that is less impactful on consumers' personal lives than a robo-advisor. On a related note, we here positioned women in the disadvantage and men in the advantage (i.e., the setting that corresponds to biases in human financial advice) but it is not certain if we were to obtain the same results if the (dis-)advantage was distributed the other way round. Future work could explore these different scenarios to help generalize and better understand the relationship between the effect of disparate impact on consumer trust.

Second, our finding that advantaged and disadvantaged users experienced the same decrease in trust appears to go counter to previous research suggesting that people make stronger fairness judgments when they are personally affected~\cite{Ham2008a}. However, it is not clear from our results to what degree advantaged users (i.e., men) felt personally affected; e.g., because they have women in their lives who they deeply care about. The role of personal relevance in the effect of disparate impact on consumer trust thus remains to be clarified by future research.

Third, our results show a decreasing trend in consumer trust as conditions become more extreme, but show a statistically significant difference only between the control and extreme bias conditions. Future work could examine these differences (also across domains) in more detail to establish the relationship between \textit{the level of} disparate impact and consumer trust (e.g., to determine what lies within and beyond an ``acceptable margin'' of disparate impact).

%NT: Can we link to previous work here, to say previous work found an effect of individual fairness perceptions on trust.
Fourth, we studied fairness related to group membership (i.e., gender), which might elicit a different (moral) evaluation than fairness on the individual level. Our results show that trust can decrease despite perceived personal benefit. However, this effect might have been caused by a sense of loyalty towards the disadvantaged group. An interesting direction for future work is to study whether similar patterns emerge when disparate impact concerns individuals; e.g., when advantaged and disadvantaged subjects are randomly chosen.

\section{Conclusion}

We presented a user study investigating the effect of algorithmically-driven disparate impact (i.e., when algorithm outcomes adversely affect one group of consumers compared to another) on consumer trust. Specifically, we studied the effect of \textit{gender-bias} in an application that aimed to persuade consumers' to make better financial decisions. We found that disparate impact decreased participants' trust and willingness to use the application. Furthermore, our results show that the trust of unfairly advantaged participants was just as affected as that of disadvantaged participants. These findings imply that disparate impact (i.e., as a result of algorithmic unfairness) can undermine trust in consumer-oriented AI systems and should therefore be avoided or mitigated when aiming to create trustworthy technology.

\section*{Acknowledgements}

This research has been supported by the \emph{Think Forward Initiative} (a partnership between ING Bank, Deloitte, Dell Technologies, Amazon Web Services, IBM, and the Center for Economic Policy Research – CEPR). The views and opinions expressed in this paper are solely those of the authors and do not necessarily reflect the official policy or position of the Think Forward Initiative or any of its partners.

%---------- Bibliography ----------%

\bibliographystyle{splncs04}
\bibliography{library.bib}

\begin{thebibliography}{10}
\providecommand{\url}[1]{\texttt{#1}}
\providecommand{\urlprefix}{URL }
\providecommand{\doi}[1]{https://doi.org/#1}

\bibitem{Ahmad2018}
Ahmad, W.N.W., Ali, N.M.: {A Study on Persuasive Technologies: The Relationship
  between User Emotions, Trust and Persuasion}. Int. J. Interact. Multimed.
  Artif. Intell.  \textbf{5}(1),  57--61 (2018).
  \doi{10.9781/ijimai.2018.02.010}

\bibitem{angwin2019machine}
Angwin, J., Larson, J., Mattu, S., Kirchner, L.: {Machine bias: There's
  software used across the country to predict future criminals and it's biased
  against blacks}. ProPublica  (2019),
  \url{https://www.propublica.org/article/machine-bias-risk-assessments-in-criminal-sentencing}

\bibitem{Araujo2020}
Araujo, T., Helberger, N., Kruikemeier, S., de~Vreese, C.H.: {In AI we trust?
  Perceptions about automated decision-making by artificial intelligence}. AI
  Soc.  \textbf{35}(3),  611--623 (2020). \doi{10.1007/s00146-019-00931-w},
  \url{https://doi.org/10.1007/s00146-019-00931-w}

\bibitem{Arnold2019}
Arnold, M., Piorkowski, D., Reimer, D., Richards, J., Tsay, J., Varshney, K.R.,
  Bellamy, R.K., Hind, M., Houde, S., Mehta, S., Mojsilovic, A., Nair, R.,
  Ramamurthy, K.N., Olteanu, A.: {FactSheets: Increasing trust in AI services
  through supplier's declarations of conformity}. IBM J. Res. Dev.
  \textbf{63}(4-5) (2019). \doi{10.1147/JRD.2019.2942288}

\bibitem{Baeckstrom2018}
Baeckstr{\"{o}}m, Y., Silvester, J., Pownall, R.A.: {Millionaire investors:
  financial advisors, attribution theory and gender differences}. Eur. J.
  Financ.  \textbf{24}(15),  1333--1349 (2018).
  \doi{10.1080/1351847X.2018.1438301}

\bibitem{Barocas2016}
{Barocas, Solon and Selbst}, A.D.: {Big data's disparate impact}. Calif. Law
  Rev.  \textbf{104}(671),  671--732 (2016)

\bibitem{Bellamy2019}
Bellamy, R.K., Mojsilovic, A., Nagar, S., Ramamurthy, K.N., Richards, J., Saha,
  D., Sattigeri, P., Singh, M., Varshney, K.R., Zhang, Y., Dey, K., Hind, M.,
  Hoffman, S.C., Houde, S., Kannan, K., Lohia, P., Martino, J., Mehta, S.: {AI
  Fairness 360: An extensible toolkit for detecting and mitigating algorithmic
  bias}. IBM J. Res. Dev.  \textbf{63}(4-5) (2019).
  \doi{10.1147/JRD.2019.2942287}

\bibitem{Corbett-Davies2018}
Corbett-Davies, S., Goel, S.: {The Measure and Mismeasure of Fairness: A
  Critical Review of Fair Machine Learning}. arXiv Prepr. arXiv1808.00023
  (2018), \url{http://arxiv.org/abs/1808.00023}

\bibitem{Cramer2016}
Cramer, A.O., van Ravenzwaaij, D., Matzke, D., Steingroever, H., Wetzels, R.,
  Grasman, R.P., Waldorp, L.J., Wagenmakers, E.J.: {Hidden multiplicity in
  exploratory multiway ANOVA: Prevalence and remedies}. Psychon. Bull. Rev.
  \textbf{23}(2),  640--647 (2016). \doi{10.3758/s13423-015-0913-5}

\bibitem{Cummings2004}
Cummings, M.L.: {Automation bias in intelligent time critical decision support
  systems}. Collect. Tech. Pap. - AIAA 1st Intell. Syst. Tech. Conf.
  \textbf{2},  557--562 (2004). \doi{10.2514/6.2004-6313}

\bibitem{Diab2011}
Diab, D.L., Pui, S.Y., Yankelevich, M., Highhouse, S.: {Lay perceptions of
  selection decision aids in US and non-US samples}. Int. J. Sel. Assess.
  \textbf{19}(2),  209--216 (jun 2011). \doi{10.1111/j.1468-2389.2011.00548.x}

\bibitem{Dietvorst2015}
Dietvorst, B.J., Simmons, J.P., Massey, C.: {Algorithm aversion: People
  erroneously avoid algorithms after seeing them err}. J. Exp. Psychol. Gen.
  \textbf{144}(1),  114--126 (2015). \doi{10.1037/xge0000033}

\bibitem{Feldman2015}
Feldman, M., Friedler, S.A., Moeller, J., Scheidegger, C., Venkatasubramanian,
  S.: {Certifying and Removing Disparate Impact}. In: Proc. 21th ACM SIGKDD
  Int. Conf. Knowl. Discov. data Min. pp. 259--268 (2015)

\bibitem{Ham2008a}
Ham, J., van~den Bos, K.: {Not fair for me! The influence of personal relevance
  on social justice inferences}. J. Exp. Soc. Psychol.  \textbf{44}(3),
  699--705 (2008). \doi{10.1016/j.jesp.2007.04.009}

\bibitem{Hardt2016}
Hardt, M., Price, E., Srebro, N.: {Equality of opportunity in supervised
  learning}. Adv. Neural Inf. Process. Syst. pp. 3323--3331 (2016)

\bibitem{JASP2020}
{JASP Team}: {JASP (Version 0.14)} (2020)

\bibitem{jeffreys1939}
Jeffreys, H.: {Theory of Probability}. Oxford University Press, Oxford (1939)

\bibitem{Lee2013}
Lee, M.D., Wagenmakers, E.J.: {Bayesian cognitive modeling: A practical
  course}. Cambridge University Press (2014). \doi{10.1017/CBO9781139087759}

\bibitem{Lee2018}
Lee, M.K.: {Understanding perception of algorithmic decisions: Fairness, trust,
  and emotion in response to algorithmic management}. Big Data Soc.
  \textbf{5}(1),  1--16 (2018). \doi{10.1177/2053951718756684}

\bibitem{Lieber2014}
Lieber, R.: {Financial Advice for People Who Aren't Rich} (apr 2014),
  \url{https://www.nytimes.com/2014/04/12/your-money/start-ups-offer-financial-advice-to-people-who-arent-rich.html}

\bibitem{Mary2019}
Mary, J.J., Calauz{\`{e}}nes, C., Karoui, N.E.: {Fairness-Aware Learning for
  Continuous Attributes and Treatments}. Icml  \textbf{97},  4382--4391 (2019),
  \url{http://proceedings.mlr.press/v97/mary19a.html}

\bibitem{Mehrabi2019}
Mehrabi, N., Morstatter, F., Saxena, N., Lerman, K., Galstyan, A.: {A Survey on
  Bias and Fairness in Machine Learning}. arXiv Prepr. arXiv1908.09635  (2019),
  \url{http://arxiv.org/abs/1908.09635}

\bibitem{Muir1996}
{Muir, Bonnie M and Moray}, N.: {Trust in automation. Part II. Experimental
  studies of trust and human intervention in a process control simulation}.
  Ergonomics  \textbf{39}(3),  429--460 (1996)

\bibitem{Mullainathan2012}
Mullainathan, S., Noeth, M., Schoar, A.: {The Market for Financial Advice: An
  Audit Study}. SSRN Electron. J.  (2012). \doi{10.2139/ssrn.1572334}

\bibitem{Napierala2012}
{Napierala, M}, A.: {What Is the Bonferroni correction?} (2012),
  \url{http://www.aaos.org/news/aaosnow/apr12/research7.asp}

\bibitem{Nickel2012}
Nickel, P., Spahn, A.: {Trust, Discourse Ethics, and Persuasive Technology}.
  In: Persuas. Technol. Des. Heal. Safety; 7th Int. Conf. Persuas. Technol.
  2012. pp. 37--40. Link{\"{o}}ping University Electronic Press (2012)

\bibitem{Ntoutsi2020}
Ntoutsi, E., Fafalios, P., Gadiraju, U., Iosifidis, V., Nejdl, W., Vidal, M.E.,
  Ruggieri, S., Turini, F., Papadopoulos, S., Krasanakis, E., Kompatsiaris, I.,
  Kinder-Kurlanda, K., Wagner, C., Karimi, F., Fernandez, M., Alani, H.,
  Berendt, B., Kruegel, T., Heinze, C., Broelemann, K., Kasneci, G., Tiropanis,
  T., Staab, S.: {Bias in data-driven artificial intelligence systems—An
  introductory survey}. Wiley Interdiscip. Rev. Data Min. Knowl. Discov.
  \textbf{10}(3),  1--14 (2020). \doi{10.1002/widm.1356}

\bibitem{Oinas-Kukkonen2008}
Oinas-Kukkonen, H., Harjumaa, M.: {Towards deeper understanding of persuasion
  in software and information systems}. In: Proc. 1st Int. Conf. Adv. Comput.
  Interact. ACHI 2008 (2008). \doi{10.1109/ACHI.2008.31}

\bibitem{Onkal2009}
{\"{O}}nkal, D., Goodwin, P., Thomson, M., G{\"{o}}n{\"{u}}l, S., Pollock, A.:
  {The relative influence of advice from human experts and statistical methods
  on forecast adjustments}. J. Behav. Decis. Mak.  \textbf{22}(4),  390--409
  (2009). \doi{10.1002/bdm.637}

\bibitem{Orji2018}
Orji, R., Moffatt, K.: {Persuasive technology for health and wellness:
  State-of-the-art and emerging trends}. Health Informatics J.  \textbf{24}(1),
   66--91 (2018). \doi{10.1177/1460458216650979}

\bibitem{Otterbacher2018}
Otterbacher, J., Checco, A., Demartini, G., Clough, P.: {Investigating user
  perception of gender bias in image search: The role of sexism}. 41st Int. ACM
  SIGIR Conf. Res. Dev. Inf. Retrieval, SIGIR 2018 pp. 933--936 (2018).
  \doi{10.1145/3209978.3210094}

\bibitem{Promberger2006}
Promberger, M., Baron, J.: {Do patients trust computers?} J. Behav. Decis. Mak.
   \textbf{19}(5),  455--468 (2006). \doi{10.1002/bdm.542}

\bibitem{Purpura2011}
Purpura, S., Schwanda, V., Williams, K., Stubler, W., Sengers, P.: {Fit4Life:
  The Design of a Persuasive Technology Promoting Healthy Behavior and Ideal
  Weight}. In: Proc. SIGCHI Conf. Hum. factors Comput. Syst. pp. 423--432
  (2011)

\bibitem{Rossi2019}
Rossi, F.: {Building trust in artificial intelligence}. J. Int. Aff.
  \textbf{72}(1),  127--133 (2019)

\bibitem{Sattarov2019}
Sattarov, F., Nagel, S.: {Building trust in persuasive gerontechnology:
  User-centric and institution-centric approaches}. Gerontechnology
  \textbf{18}(1),  1--14 (2019). \doi{10.4017/gt.2019.18.1.001.00}

\bibitem{Smith2020}
Smith, J., Sonboli, N., Fiesler, C., Burke, R.: {Exploring User Opinions of
  Fairness in Recommender Systems}. In: CHI'20 Work. Human-Centered Approaches
  to Fair Responsible AI (2020), \url{http://arxiv.org/abs/2003.06461}

\bibitem{Toreini2020}
Toreini, E., Aitken, M., Coopamootoo, K., Elliott, K., Zelaya, C.G., van
  Moorsel, A.: {The relationship between trust in AI and trustworthy machine
  learning technologies}. FAT* 2020 - Proc. 2020 Conf. Fairness,
  Accountability, Transpar. pp. 272--283 (2020). \doi{10.1145/3351095.3372834}

\bibitem{VanDenBergh2020}
{Van Den Bergh}, D., {Van Doorn}, J., Marsman, M., Draws, T., {Van Kesteren},
  E.J., Derks, K., Dablander, F., Gronau, Q.F., Kucharsk{\'{y}}, {\v{S}}.,
  Gupta, A.R.N., Sarafoglou, A., Voelkel, J.G., Stefan, A., Ly, A., Hinne, M.,
  Matzke, D., Wagenmakers, E.J.: {A tutorial on conducting and interpreting a
  bayesian ANOVA in JASP}. Annee Psychol.  \textbf{120}(1),  73--96 (2020).
  \doi{10.3917/anpsy1.201.0073}

\bibitem{Varshney2019}
Varshney, K.R.: {Trustworthy machine learning and artificial intelligence}.
  XRDS Crossroads, ACM Mag. Students  \textbf{25}(3) (2019).
  \doi{10.1145/3313109}

\bibitem{Varshney2020}
Varshney, K.R.: {On Mismatched Detection and Safe, Trustworthy Machine
  Learning}. In: 2020 54th Annu. Conf. Inf. Sci. Syst. CISS 2020 (2020).
  \doi{10.1109/CISS48834.2020.1570627767}

\bibitem{Verbeek2006}
Verbeek, P.P.: {Persuasive Technology and Moral Responsibility Toward an
  ethical framework for persuasive technologies}. Persuasive  \textbf{6},
  1--15 (2006)

\bibitem{Verma2018}
Verma, S., Rubin, J.: {Fairness definitions explained}. In: Proc. Int. Work.
  Softw. Fairness. pp.~1--7. FairWare '18, Association for Computing Machinery,
  New York, NY, USA (2018). \doi{10.1145/3194770.3194776}

\bibitem{vigdor2019apple}
Vigdor, N.: {Apple card investigated after gender discrimination complaints}.
  New York Times  (2019)

\bibitem{Wagenmakers2018}
Wagenmakers, E.J., Marsman, M., Jamil, T., Ly, A., Verhagen, J., Love, J.,
  Selker, R., Gronau, Q.F., {\v{S}}m{\'{i}}ra, M., Epskamp, S., Matzke, D.,
  Rouder, J.N., Morey, R.D.: {Bayesian inference for psychology. Part I:
  Theoretical advantages and practical ramifications}. Psychon. Bull. Rev.
  \textbf{25}(1),  35--57 (2018). \doi{10.3758/s13423-017-1343-3}

\bibitem{Woodruff2018}
Woodruff, A., Fox, S.E., Rousso-Schindler, S., Warshaw, J.: {A qualitative
  exploration of perceptions of algorithmic fairness}. Conf. Hum. Factors
  Comput. Syst. - Proc.  \textbf{2018-April},  1--14 (2018).
  \doi{10.1145/3173574.3174230}

\bibitem{Yang2018a}
Yang, Q., Banovic, N., Zimmerman, J.: {Mapping machine learning advances from
  HCI research to reveal starting places for design innovation}. Conf. Hum.
  Factors Comput. Syst. - Proc.  \textbf{2018-April},  1--11 (2018).
  \doi{10.1145/3173574.3173704}

\bibitem{Zafar2017}
Zafar, M.B., Valera, I., Rodriguez, M.G., Gummadi, K.P.: {Fairness beyond
  disparate treatment {\&} disparate impact: Learning classification without
  disparate mistreatment}. 26th Int. World Wide Web Conf. WWW 2017 pp.
  1171--1180 (2017). \doi{10.1145/3038912.3052660}

\end{thebibliography}

\end{document}